\documentclass[10pt,conference]{IEEEtran}
\IEEEoverridecommandlockouts
\usepackage{cite}
\usepackage{amsmath,amssymb,amsfonts}
\usepackage{algorithm}
\usepackage{algorithmicx}
\usepackage{algpseudocode}

\usepackage{graphicx}
\usepackage{balance}
\usepackage{textcomp}
\usepackage{xcolor}
\usepackage{url}
\def\BibTeX{{\rm B\kern-.05em{\sc i\kern-.025em b}\kern-.08em
    T\kern-.1667em\lower.7ex\hbox{E}\kern-.125emX}}

\usepackage[most]{tcolorbox}

\usepackage{url}

\usepackage{breakurl}

\newcommand{\tool}{\textsc{Chronos}}

\begin{document}

\author{
    \IEEEauthorblockN{Yunbo Lyu\IEEEauthorrefmark{1}\textsuperscript{\textsection}, Thanh Le-Cong\IEEEauthorrefmark{1}\textsuperscript{\textsection}, Hong Jin Kang\IEEEauthorrefmark{1}, Ratnadira Widyasari\IEEEauthorrefmark{1},  \\ Zhipeng Zhao\IEEEauthorrefmark{1},  Xuan-Bach D. Le\IEEEauthorrefmark{2},  Ming Li\IEEEauthorrefmark{3}, David Lo\IEEEauthorrefmark{1}}
    
    \IEEEauthorblockA{\IEEEauthorrefmark{1}Singapore Management University  \IEEEauthorrefmark{2}The University of Melbourne
    \IEEEauthorrefmark{3}Nanjing University} 
    
    \{yunbolyu, tlecong\}@smu.edu.sg, \{hjkang.2018, ratnadiraw.2020\}@phdcs.smu.edu.sg, zpzhao@smu.edu.sg, \\
    bach.le@unimelb.edu.au, lim@nju.edu.cn, davidlo@smu.edu.sg
}

\title{\tool{}: Time-Aware Zero-Shot Identification of Libraries from Vulnerability Reports\\
}



\maketitle
\begingroup\renewcommand\thefootnote{\textsection}
\footnotetext{Equal contribution}
\endgroup
\begin{abstract}


Tools that alert developers about library vulnerabilities depend on accurate, up-to-date vulnerability databases which are maintained by security researchers.
These databases record the libraries related to each vulnerability. However, the vulnerability reports may not explicitly list every library and human analysis is required to determine all the relevant libraries.
Human analysis may be slow and expensive, which motivates the need for automated approaches. 
Researchers and practitioners have proposed to automatically identify libraries from vulnerability reports using extreme multi-label learning (XML).

While state-of-the-art XML techniques showed promising performance, their experimental settings do not practically fit what happens in reality. 
Previous studies randomly split the vulnerability reports data for training and testing their models without considering the chronological order of the reports. 
This may unduly train the models on chronologically newer reports while testing the models on chronologically older ones. 
However, in practice, one often receives chronologically new reports, which may be related to previously unseen libraries. 
Under this practical setting, we observe that the performance of current XML techniques declines substantially, e.g., F1 decreased from 0.7 to {\color{black}0.28} under experiments without and with consideration of chronological order of vulnerability reports.  



We propose a practical library identification approach, namely \tool{}, based on zero-shot learning.
The novelty of \tool{} is three-fold. First, \tool{} fits into the practical pipeline by considering the chronological order of vulnerability reports. 
Second, \tool{} enriches the data of the vulnerability descriptions and labels using a carefully designed data enhancement step. 
Third, \tool{} exploits the temporal ordering of the vulnerability reports using a cache to prioritize prediction of versions of libraries that recently had reports of vulnerabilities.  


In our experiments, \tool{} achieves an average F1-score of 0.75, {\color{black}2.7x} 
better than the best XML-based approach.
Data enhancement and the time-aware adjustment improve \tool{} over the vanilla zero-shot learning model by 27\% in average F1.

\end{abstract}

\begin{IEEEkeywords}
zero-shot learning, library identification, unseen labels, extreme multi-label classification, vulnerability reports
\end{IEEEkeywords}

\section{Introduction}

\begin{figure}
    \centering
    \includegraphics[width=0.8\linewidth]{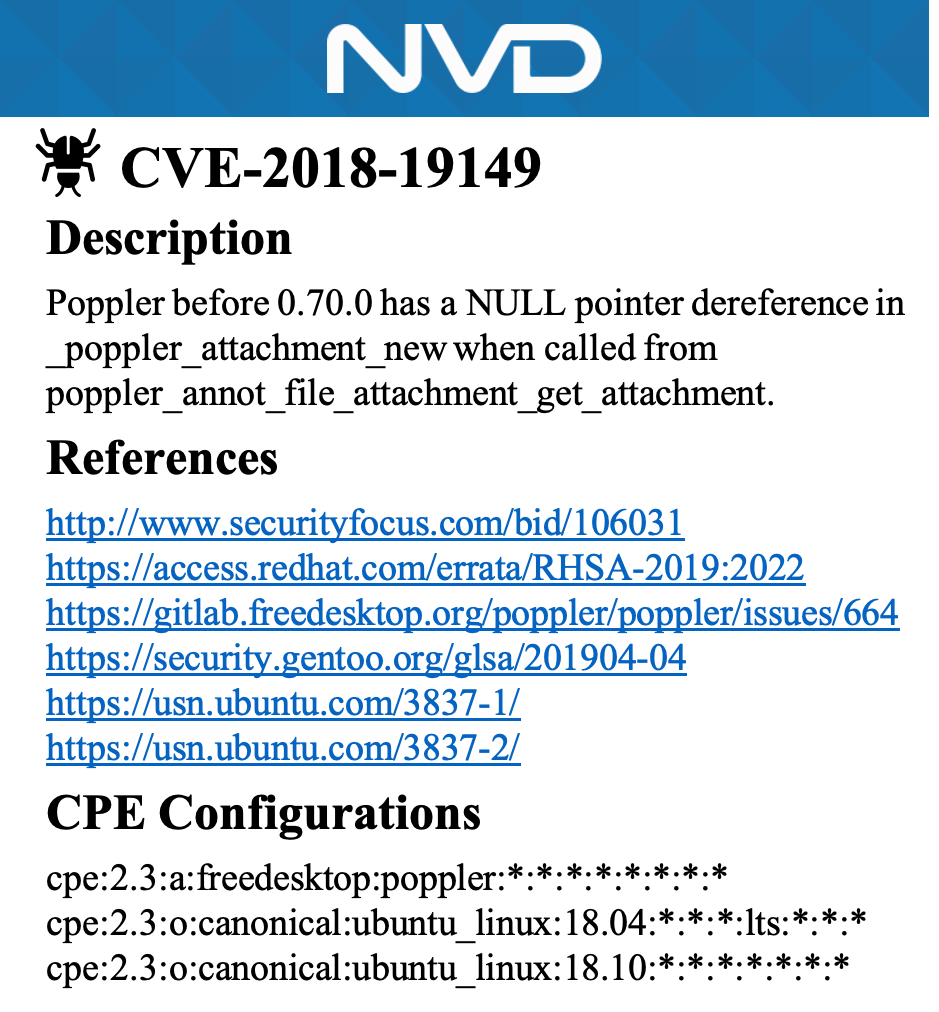}
    \caption{NVD entry for CVE-2018-19149. Each vulnerability report has a description, some references, and CPE configurations. While  ``evince'' is affected by the vulnerability, the term ``evince'' does not appear in the report.}
    \label{fig:nvd}
\end{figure}

The use of third-party libraries is commonplace in software development,
however,
software engineers have to be aware of and manage library vulnerabilities~\cite{prana2021out,imtiaz2022open,imtiaz2021comparative}.
Software Composition Analysis tools have been proposed to assist developers by warning them of vulnerable libraries included in a software project's dependencies.
These tools, including those built by industrial companies, such as Veracode~\cite{foo2019dynamics} and Snyk~\cite{snyk_monitoring}, are now widely deployed but depend on an up-to-date and accurate vulnerability database.
These databases indicate libraries, known vulnerabilities, vulnerable library versions, and other data~\cite{foo2019dynamics}.
The database is maintained by security researchers who, through their domain knowledge and manual effort, curate vulnerability reports from multiple
sources, including the National Vulnerability Database (NVD). 
A vulnerability report has an identification number,
a CVE (Common Vulnerability Enumeration) ID and a description of
the vulnerability. 
While a CPE (Common Platform Enumeration) configuration indicates a 
package or library that is related to the vulnerability, this configuration is not exhaustive~\cite{chen2020automated}.
Hence, security researchers have to annotate each vulnerability report with the affected libraries and even specific versions if they think the versions are noteworthy.
For alerting developers, these databases require a mapping between each vulnerability ID and the libraries (and specific versions) that are affected by the vulnerability.
For example, Figure~\ref{fig:nvd} shows the vulnerability report of 
CVE-2018-19149. 
While the vulnerability report mentions ``Poppler'', other software systems such as ``evince'' and ``okular'' are also affected~\cite{snyk_cve}.


There is usually a delay from vulnerability disclosure (i.e., publicly posted on NVD and assigned a CVE ID)  to developers updating their dependencies~\cite{kula2018developers,chinthanet2021lags}.
This motivates automated approaches that speed up the work of security researchers.
For curating vulnerabilities to update vulnerability databases, Chen et al. from Veracode
have proposed to automatically identify libraries from the vulnerability reports~\cite{chen2020automated}.
The study has formulated the problem as an extreme multi-label classification (XML) problem~\cite{chen2020automated}. Characterized by the sparsity  of the data and the large space of possible labels, XML problems are challenging for standard machine learning approaches.
Recently, Haryono et al.~\cite{stefanus2022automated} found that the most effective XML approach for library identification is a deep learning-based approach, LightXML~\cite{jiang2021lightxml}. 

While XML techniques were shown to be effective in the experiments of prior studies~\cite{chen2020automated,stefanus2022automated}, we observe that there are practical concerns that need to be addressed.
Every year, new libraries are included in the NVD. 
If an XML approach is trained strictly on data prior to the inclusion of the new library, it would not produce the correct labels as output.
In other words, existing library identification approaches will fail to predict previously unseen libraries.

We performed an empirical study of the number of new libraries with vulnerabilities each year.
Our analysis indicates that up to 70\% libraries associated up to 50.7\% of vulnerability reports each year cannot be correctly identified by the previously proposed approaches~\cite{chen2020automated,stefanus2022automated}.
As the training dataset would not contain any NVD entries related to the libraries, the XML techniques would not correctly identify vulnerabilities related to these libraries.
To address this practical concern, we reformulate the library identification task as a generalized zero-shot learning task.
We split the dataset chronologically; as new libraries may be added to the NVD, there would be libraries in the testing dataset that do not correspond to any NVD entry in the training dataset.


To tackle the aforementioned task, we propose a practical library identification approach namely \tool{} based on zero-shot XML, that is capable of predicting previously unseen labels from vulnerability reports. 
To achieve this, \tool{} relies on two main observations: (1) Additional documents referenced in the NVD entry, e.g., bug reports, mailing lists, can help distinguish multiple previously unseen labels from one another. 
(2) Exploiting temporal connection between vulnerability reports and affected libraries can help boost the prediction accuracy. 
The key intuition is that if a vulnerability was reported for a particular version of a library recently, it is likely that new vulnerabilities will be reported for the same version rather than for an older version. 

Towards this end, \tool{} implements several techniques to retrieve and process additional sources of information referenced from NVD entries, enriching the vulnerability description with more information. 
To exploit temporal connection between vulnerability reports and affected libraries, \tool{} uses a cache to track the libraries related to the most recently seen NVD entries. 
The cache enables a reranking of \tool{}'s predictions by favouring libraries and versions that were most recently observed to be vulnerable.

In our experiments, \tool{} achieves an average F1 of 0.75, outperforming the LightXML \cite{jiang2021lightxml}   
approach by {\color{black}167.9\% }in average F1 (0.75 to {\color{black}0.28}).
This demonstrates the superior performance of \tool{} over traditional non zero-shot XML models.
Compared to a manually handcrafted approach that directly matches library names
against the vulnerability description, \tool{} performs 92.3\% better.

Compared to an approach using only the CPE, \tool{} performs 3 times better.
Our analysis reveals that each component of \tool{} contributes positively to its effectiveness. 
Removing the data enhancement step reduces performance by 6.7\%.
Removing the time-aware adjustment reduces performance by 9.2\%. Overall, \tool{} improves over a vanilla zero-shot XML model by 27\% (0.75 vs. 0.59).

Our study has practical and research significance. 
It helps in securing the software supply chain
by automating slow manual analysis, and
highlights practical concerns, such as 
the chronological order of data, in developing automated tools.

In summary, our paper makes the following contributions:

\begin{itemize}
    \item \textbf{Problem reformulation.} We reformulate the task of predicting libraries to consider the reports chronologically based on publication dates.
    The task is a generalized zero-shot extreme multi-label (XML) classification task; vulnerability reports in the testing dataset may be related to libraries that did not appear in the training dataset.
    \item \textbf{Approach.} We propose  \tool{}, a zero-shot learning technique. \tool{} uses data enrichment and a time-aware adjustment step to favour more recently seen versions of each library.
    \tool{}'s dataset \cite{dataset} and implementation \cite{implementation} are publicly available.
    
    \item \textbf{Experiments.} We evaluate \tool{} and show that  \tool{} outperforms the strongest previously proposed approach by {\color{black}167.9\%} in average F1 on the realistic but more challenging experimental setting. 
\end{itemize}

The rest of this paper is organized as follows. 
Section \ref{sec:background} covers the background of our work. 
Section \ref{sec:formulation} formulates the task. 
Section \ref{sec:approach} introduces \tool{}. 
Section \ref{sec:experiments} discusses our experimental results. 
Section \ref{sec:discussion} presents a deeper analysis of our findings and threats to validity. 
Finally, Section \ref{sec:conclusion} concludes the paper.

\section{Background}
\label{sec:background}


\subsection{Extreme Multi-label Classification for Identifying Libraries}

Extreme Multi-label Learning (XML) models 
assign relevant labels to documents~\cite{prabhu2014fastxml,jiang2021lightxml}. 
Each document may be assigned multiple labels.
Tasks employing XML techniques are characterized by an extremely large label space and sparse data. 
XML approaches have to select a small subset of relevant labels out of \textit{millions of possible labels}. Moreover, many labels have only a few instances associated with them, posing a challenge for standard machine learning techniques~\cite{prabhu2014fastxml,jiang2021lightxml}.

Chen et al.~\cite{chen2020automated} from Veracode, a well-known application security company, formulated the task of identifying libraries affected by vulnerabilities given the vulnerability report.
Their experiments revealed that the NVD report's CVE configuration was insufficient for identifying \textit{every} affected library. 
Their experiments highlighted the promise of applying XML techniques for the task.
Each vulnerability report may describe multiple affected libraries, and the space of all libraries is enormous.
These characteristics present challenges for traditional Machine Learning techniques but are addressed by XML techniques.
A recent study by Haryono et al.~\cite{stefanus2022automated} assessed recent XML techniques on the task.
Their experiments revealed that the deep learning-based approach, LightXML~\cite{jiang2021lightxml}, led to the greatest increase in performance among recently proposed XML techniques.

We show that while the powerful XML techniques had strong performance in the experiments of prior studies, the experiments did not  capture every practical consideration.
In this study, we reformulate the task as a generalized zero-shot learning problem.

\subsection{Generalized Zero-Shot Learning}

The challenge of predicting labels that do not appear during training is established in the machine learning literature~\cite{wang2019survey}. 
In zero-shot learning, the training and testing labels are disjoint.
In generalized zero-shot learning, both seen and unseen labels appear in the testing dataset.

For generalized zero-shot XML problems, ZestXML~\cite{gupta2021generalized} has been previously proposed.
ZestXML aims to exploit the sparsity of the data in XML tasks.
During training, ZestXML learns to project a
small number of features to be close to the features of the relevant labels.
Using a novel optimization technique based on the assumption that only a few features are relevant to a label, ZestXML is able to be trained quickly.
We use ZestXML in our approach
as it is targeted at zero-shot learning tasks; ZestXML can output labels without any training data as long as the document features closely match the label features (c.f., Section~\ref{subsec:zestxml}).

\section{Problem Formulation}
\label{sec:formulation}

\subsection{Usage Scenario}
\label{subsec:usage}


A security researcher is monitoring and curating vulnerability data from multiple sources, including the National Vulnerability Database (NVD).
For each vulnerability report, the researcher has to map it to a set of relevant libraries.
Without an automated tool, the security researcher has to rely only on their domain knowledge and carefully analyze the vulnerability description and references to mailing lists/bug reports. 
Unfortunately, there is a large number of vulnerability reports and many possible libraries, and even each libraries may have many different versions. 
As a result, human analysis is slow and may be error-prone.
An automated approach that predicts relevant libraries would augment the manual analysis performed by the researcher.

\subsection{Problem Formulation}
\label{subsec:formulation}

In this work, following prior works \cite{chen2020automated, stefanus2022automated}, we formulate the problem of library identification from vulnerability reports as an XML problem, where vulnerability reports and possible libraries, which can be enumerated from package managers 
(e.g., npm and pypi), are considered documents to be classified and their labels, respectively.
If security researchers believe that particular versions of the libraries are noteworthy\cite{chen2020automated}, the vulnerability report may be labelled with specific versions of the affected library (e.g. the standard library of java 1.7 vs java 1.8). 
Different from prior works, we reformulate the problem in the zero-shot setting as follows. 

\vspace{2mm}

\noindent \textbf{Prior Knowledge.} A labelled dataset $\mathcal{D} = (\mathcal{V}, \mathcal{L}, \mathcal{M})$, where $\mathcal{V}$ is set of vulnerability reports and $\mathcal{L}$ is set of labels for $\mathcal{V}$ is a mapping from $\mathcal{V}$ to set of subsets of $\mathcal{L}$, where $\mathcal{M}(v) \subseteq \mathcal{L}$ is the set of labels for a vulnerability report $v$. A label $l \in \mathcal{L}$, may identify a particular library version.

\vspace{2mm}

\noindent \textbf{Input.}  A new (unlabelled) dataset $\mathcal{D}_{new} = (\mathcal{V}_{new}, \mathcal{L}_{new})$ where $\mathcal{V}_{new} \neq \mathcal{V}$ is set of new vulnerability reports and $\mathcal{L}_{new} \supseteq \mathcal{L}$ is set of labels. 

\vspace{2mm}

\noindent \textbf{Output.} A mapping $\mathcal{M}_{new}$ from $\mathcal{V}_{new}$ to set of subsets of $\mathcal{L}_{new}$ such that $\mathcal{M}_{new}(v) \subseteq \mathcal{L}_{new}$ is the set of labels for each vulnerability report $v \in \mathcal{V}_{new}$. 

\vspace{2mm}

The approaches proposed in prior studies~\cite{stefanus2022automated, chen2020automated} are built upon the assumption that the set of labels in the labelled (training) dataset $\mathcal{L}_{new}$ are identical to the labels in new (testing) dataset $\mathcal{L}$. More formally, 
\begin{equation}
    \label{eq:sup_assume}
   \mathcal{L}_{new} = \mathcal{L}
\end{equation}
The previous studies treat the problem as a supervised learning task. As a result, they trained supervised learning models such as LightXML~\cite{jiang2021lightxml} or FastXML~\cite{prabhu2014fastxml} to learn the mapping $\mathcal{M}$ from labelled dataset and then use the trained model as the mapping $\mathcal{M}_{new}$. 
Unfortunately, in a practical setting, the assumption (\ref{eq:sup_assume}) does not hold.


In this paper, we reformulate the problem of library identification from vulnerability reports to consider the possibility of unseen libraries. 
We relax assumption (\ref{eq:sup_assume}) to:
\begin{equation}
    \mathcal{L}_{new} \supseteq \mathcal{L}
\end{equation}
This assumption means that $\mathcal{L}$, i.e., the set of \textit{seen labels} belonging to the labelled dataset, should be a subset of $\mathcal{L}_{new}$, i.e., the set of \textit{all seen and unseen labels}. 
This relaxation allows our problem formulation to include unseen labels and be more suitable in practical settings. 




\section{Proposed Approach}
\label{sec:approach}

\begin{figure}
    \centering
    \includegraphics[width=0.9\columnwidth]{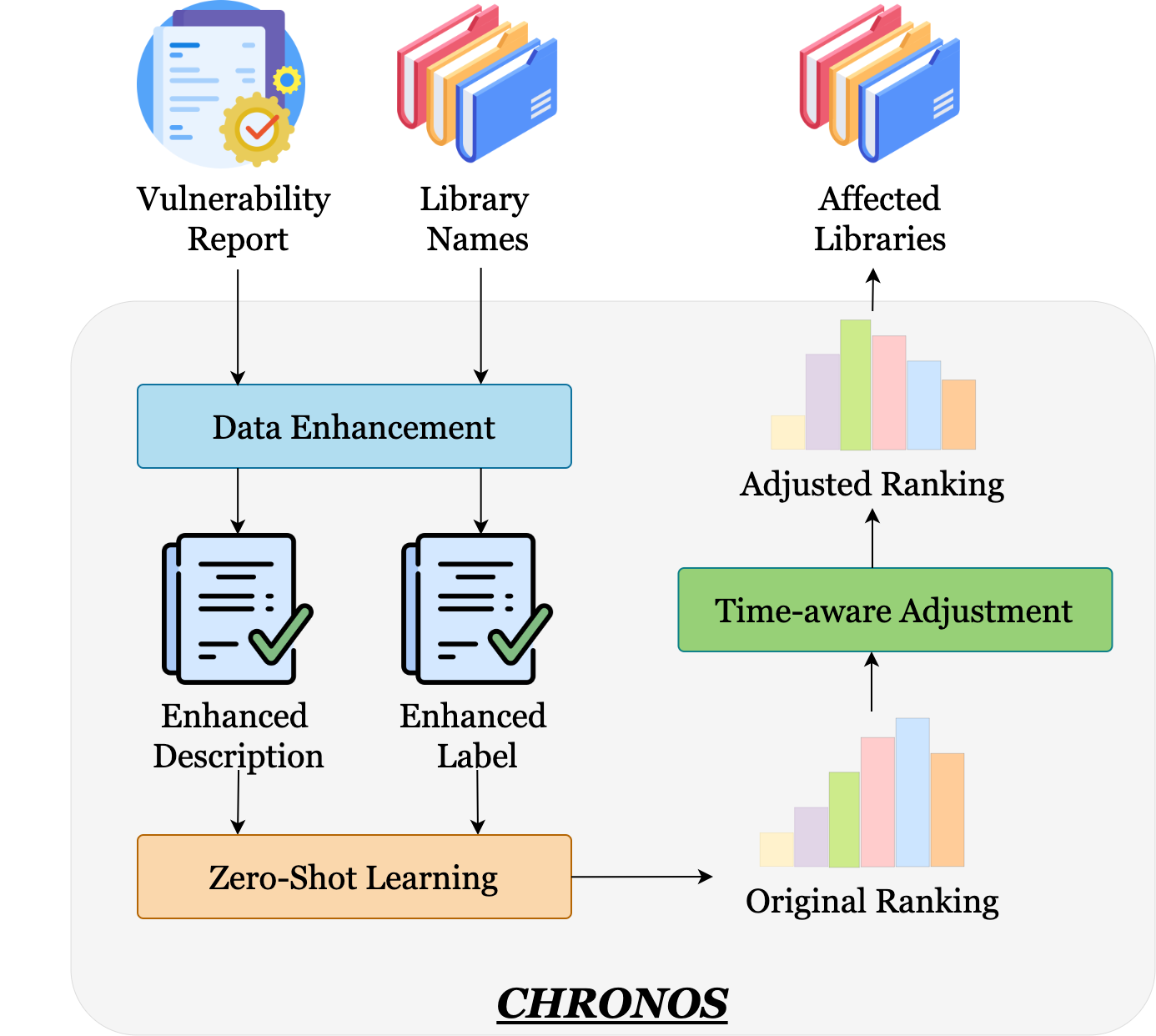}
    \caption{The overview of \tool{} }
    \label{fig:overview}
\end{figure}

Figure \ref{fig:overview} illustrates the overall framework of \tool{}. 
\tool{} identifies libraries for each vulnerability report.
There are three main components in \tool{}: (1) data enhancement, (2) a zero-shot learning XML model, and (3) time-aware adjustment. 

The first component, data enhancement (Section \ref{subsec:description}), 
addresses the lack of discriminating information to identify links between the vulnerability description and possible labels.
This enriches vulnerability descriptions by collecting data from their website references (Section \ref{subsec:collect}) and then performing preprocessing (Section \ref{subsec:preprocess}) to clean the descriptions. 
\tool{} also enriches the label features by splitting labels to sub-words (section \ref{subsec:label}).
The second component, a zero-shot XML model (Section \ref{subsec:zestxml}), uses a ZestXML model to answer the question, ``How likely is a vulnerability description and a library name to be relevant?'' using the features from vulnerability descriptions and the label features. 
The third component, time-aware adjustment (Section \ref{subsec:rerank}), 
exploits the temporal
ordering of the vulnerability reports using a cache to prioritize
predictions of versions of libraries that were more recently affected by vulnerabilities. 


\subsection{Data Enhancement}
\label{subsec:description}

\subsubsection{Collecting Reference Data}
\label{subsec:collect}

A vulnerability report can come with a list of website references. 
These website references can include any pertinent references (e.g., solutions, workarounds, etc). 
This information may be helpful to identify affected libraries. 
Before a vulnerability report is passed to the zero-shot XML model, \tool{} fetches the web references from the vulnerability report to extract the aforementioned textual information.


Unfortunately, there are many different domains (e.g. 1,054 unique domains on our dataset) for the web references.
This is a challenge for web scrapping tools as each domain has a unique web page layout. 
To address this problem, we only extract data from the highly frequent domains in our dataset as shown in Table~\ref{tab:domain}, which cover 82.3\% of the vulnerability reports in the dataset. 
For each web reference, \tool{} automatically crawls  references of depth 1, i.e., references explicitly linked from the report. 
In total, we crawled 28,783 references. 
As the quality of data collected is more important than its quantity, our crawler accommodates the different website structures of each domain to fetch the vulnerabilities’ details. 
We crawl each reference’s page title and description of the vulnerability, which may name affected products (e.g. CVE-2014-1512 references https://ubuntu.com/security/notices/USN-2151-1, which indicates that ``Thunderbird'' is affected).
\subsubsection{Preprocessing Reference Data}
\label{subsec:preprocess}
We preprocess the documents collected in the previous step in order to clean the data before using it for training \tool{}'s model. We perform the following three steps:

\begin{table}[]
\centering
\caption{Most frequent domains with their number of occurrences}
\vspace{-0.2cm}
\label{tab:domain}
\begin{tabular}{lc}
\hline
\multicolumn{1}{c}{\textbf{Domain}} & \multicolumn{1}{c}{\textbf{\#Occurrences}} \\ \hline
access.redhat.com                     & 6383                                       \\ 
list.opensuse.org                     & 4882                                       \\ 
github.com                            & 3479                                       \\ 
debian.org                            & 2853                                       \\ 
oracle.com                            & 2717                                       \\ 
securitytracker.com                   & 2286                                       \\ 
security.gentoo.org                   & 1875                                       \\ 
ubuntu.com                            & 1752                                       \\ 
usn.ubuntu.com                        & 1687                                       \\ 
openwall.com                          & 1517                                       \\ 
lists.fedoraproject.org               & 1299                                       \\ 
bugzilla.redhat.com                   & 1225                                       \\ \hline
\end{tabular}
\vspace{-0.4cm}
\end{table}

\begin{itemize}
    \item In the first \textit{basic} preprocessing step, we remove non-alphanumeric characters. 
    This preprocessing step is done using the regular expression: \textit{"[a-zA-Z][a-z]+"}. 
    \item We perform stemming and stopwords removal~\cite{yang2021incbl}.
    This is performed automatically using the spaCy~\cite{spacy} package. 
    \item We remove $x\%$ of words sorted by the number of occurrences in reference data since they are common words that will increase the noise. We also remove words that appear more than $y$ times in a single reference. 
    Some words
    appear frequently in the collected reference data.
    As they are not specific to a particular CVE or libraries, they can be removed. 
    Removing these words was previously found to be effective for identifying libraries~\cite{chen2020automated}. 
    $x$ and $y$ are parameters that are tuned on the validation dataset.
\end{itemize}

Finally, \tool{} merges the processed reference data with each vulnerability description to produce the final description, $d$, from the vulnerability reports.

\subsubsection{Library Sub-word Splitting}
\label{subsec:label}
Next, the label processing component of \tool{} enriches the features that help determine labels associated with vulnerability reports, i.e., libraries, denoted as $\mathcal{L}$.
To maximize the chance of matching a label against mentions of the library in the vulnerability description, \tool{} initializes the label features with different forms of the library name.
Library names usually comprises several words. 
While the words in some library names are visually separated based on existing conventions to ease reading (e.g. org.apache.tika),
not all libraries have names with clear conventions for splitting them (e.g. \underline{org}.\textbf{spring}\underline{framework}, \underline{py}\textbf{openssl}).
Moreover, when considering libraries related to vulnerabilities, the library names adhere to different conventions as the libraries come from a variety of different languages and ecosystems~\cite{chen2020automated}. 
Therefore, we split each library name into its constituent subtokens using a suitable library name splitter.

The decision of whether to decompose the library names into  subtokens, i.e. morphological units, has important implications. 
With subtokens, the number of features increase, making the selection the most important features of each label more difficult. 
On the other hand, breaking up a library name into multiple units has advantages as \tool{} may be able to identify more valuable features. 
Subtokens are more common than the original library name, enabling \tool{} to find more connections between unseen and seen library names, which improves \tool{} to predict unseen labels.

For splitting tokens into subtokens, several approaches use Mining Software Repositories techniques, such as LINSEN~\cite{corazza2012linsen}, Samurai~\cite{enslen2009mining}, Spiral~\cite{hucka2018spiral}. 
We apply the state-of-the-art Spiral token splitter~\cite{hucka2018spiral},
which was shown to be more effective than other methods~\cite{hucka2018spiral}. 

\subsection{Zero-shot Learning}
\label{subsec:zestxml}


The next component in \tool{} is a zero-shot learning model, which takes the vulnerability reports as input and produces a list of labels and their intermediate relevance scores as output.
Particularly, \tool{} employs the current state-of-the-art technique, ZestXML\cite{gupta2021generalized}, as the core machine learning model.
Following the original paper~\cite{gupta2021generalized}, \tool{} uses TF-IDF \cite{ramos2003using} to extract the feature vectors for the descriptions and labels. 
ZestXML takes input as the extracted features models the relevance between descriptions and labels by analyzing their linear feature interactions.
Particularly, given a description $d$ and a label $l$, the relevance score between  are calculated as follows.
\begin{equation}
    R(d,l) = d^{\top}\mathbf{W}l
    \label{eq:relevant}
\end{equation}
where a large (small) $R(d,l)$ value means high (low) relevance between $d$ and $l$. $d^{\top}$ is the transpose vector of $d$. For simplicity, ZestXML are absorbed into Equation (\ref{eq:relevant}) by appending a constant feature to $d$ and $l$. 
 
$\mathbf{W}$ is $\mathbb{R}^{D'} \times \mathbb{R}^{L'}$ a matrix of model parameters, which is learned to correctly classify all description and label pairs in the training dataset ($D'$ and $L'$ indicate their high dimensional, sparse TF-IDF feature vectors of descriptions and labels). Particularly, ZestXML learns the model parameters $W$ via a regularized logistic regression as follows.  
\begin{equation}
\begin{aligned}
&\min _{\mathbf{W}} \frac{1}{2}\|\mathbf{W}\|_{L}^{2}+\lambda \sum_{i=1}^{D} \sum_{j=1}^{L} \log \left(1+e^{-y_{ij} d_{i}^{\top} \mathbf{W}l_{j}}\right) \\
&\text { s.t., }\left\|\mathbf{W}_{i *}\right\|_{0} \leq K \text { }\forall i \in\{1, \cdots, D'\}
\end{aligned}
\label{eq:loss}
\end{equation}
where D, L are the number of descriptions and labels. $K, \lambda$ are hyper-parameters of the model and $\left\|\mathbf{W}_{i *}\right\|_{0}$ is the number of non-zeros in the $i$ th row of $\mathbf{W}$. 
$y_{ij}$ is the ground truth relevance between the description $i$ and the label $j$, where $y_{ij} \in \{-1,1\}$ and $y_{ij}=1$ denotes that description $i$ is relevant to the label $j$.

To determine the optimal model parameters $W$ for the aforementioned training objective, ZestXML proposed an extension of Hard Thresholding Pursuit~\cite{gupta2021generalized} termed XHTP. 
ZestXML was designed with the assumption that the features of each label is sparse.
Similar to other second-order optimization algorithms, an iteration of XHTP consists of 2 successive steps: (1) approximation in which XHTP approximates the training objective in Equation (\ref{eq:loss}) by a quadratic form and minimizes it to obtain a sparsified solution, and (2) refinement, in which XHTP refines the values of non-zero parameters to fit the original objective better. 
However, unlike other second-order optimization algorithms, XHTP exploits assumptions such as feature independence and a favorable starting point, i.e., $\mathbf{W} = 0$ to achieve highly sparsified and accurate model parameters in just one iteration of approximation and refinement steps. 
In this way, ZestXML improves its efficiency.

\subsection{Time-aware Adjustment}
\label{subsec:rerank}

		

\begin{algorithm} 
    \caption{Time-aware adjustment that favours new library versions and recently observed labels} 
	\label{alg:Time-aware adjustment} 
	\begin{algorithmic}[1]
	    \Require 
	   \item[]
	    \begin{itemize}
	        \item $\mathcal{L}_{highest} \gets$ top-$i$ most relevant labels for each description
	        \item version\_store $\gets$ a map of a label to newer versions of the same library
	        \item cache $\gets$ recently seen labels
	        \item $R(d,l) \gets$ a relevance
score between a description, $d$ and a  label, $l$ 
            \item $f \gets$ an update function. Given in Equation \ref{eq:update}
	    \end{itemize}
\Function{Time-Aware Adjustment}{$\mathcal{L}_{highest}$}
\For{ $l \in  \mathcal{L}_{highest}$ }
\State \Call{FavorNewVersion}{$l$, version\_store, cache} 
\EndFor
\For{ $l \in  \mathcal{L}_{highest}$ }
\State $R(d,l) \gets f(R(d,l))$
\EndFor
\EndFunction
\end{algorithmic}
\end{algorithm}


\begin{algorithm}
    \caption{Transferring the relevance scores from old to new versions of the same library} 
	\label{alg:Cache-based library reranking} 
	\begin{algorithmic}[1]
	   \Require
	   \item[]
	   \begin{itemize}
	       \item $l$ $\gets$ a label
	       \item version\_store $\gets$ a map of a label to newer versions of the same library
	       \item cache $\gets$ recently seen labels 
	       \item $R(d,l) \gets$ a relevance
score between a description, $d$ and a label, $l$
	   \end{itemize}
\Function{FavorNewVersion}{$l$, version\_store, cache}

\For{$l^{new} \in$ version\_store[$l$]}
    \If{$l^{new} \in $ cache and $R(d,l^{new}) > R(d,l)$ }
	\State  $R(d,l^{new}) \gets max(R(d,l^{new}), R(d,l) )$ 
	\State $R(d,l) \gets 0 $ 
	\State \textbf{break} 
	\EndIf
\EndFor
\EndFunction 
	\end{algorithmic}
\end{algorithm}

Next, given the labels and their relevance scores from the zero-shot XML model, the time-aware adjustment component modifies the relevance scores.
We observe that vulnerabilities in the same time range are more likely to affect the same versions of the libraries. 
Thus, \tool{} uses a strategy to prioritize versions of libraries that have been recently affected by vulnerabilities. 
In Algorithm \ref{alg:Time-aware adjustment}, \tool{}'s time-aware adjustment component uses two steps to modify the relevance scores: favor newer library versions (lines 2--4) and add a recency bias (lines 5--7).

These two steps use a version store and a cache.
The \textbf{version store} tracks the different versions of  each library. 
Given a version of a library, the version store returns the list of labels corresponding to newer versions of the library, sorted in descending order by their versions (i.e., newest versions first).

The \textbf{cache} stores the recently affected libraries using a Least Recently Used (LRU) replacement policy with a cache size $c$. 
Chronologically, as vulnerability reports are labelled with their true labels (e.g. as a security researcher annotates the ground-truth on each report after considering the predictions of \tool{}), \tool{} adds the label into the cache.
When a new label is added while the cache is full, the new label replaces the oldest entry in the cache.

For each description $d$, ZestXML models ranks the labels $l$ in set of library $\mathcal{L}_{highest}$ via the relevance scores $R(d,l)$.
\tool{} uses the cache to modify the $R(d,l)$ at prediction time through two successive steps: replacement and update.
For efficiency, \tool{} considers only the top-$i$ highest relevant labels. 
$i$ is a parameter which is tuned on the validation dataset and  will be discussed in Section \ref{subsec:implementation_details}. 

The time-aware adjustment \textbf{favors newer library versions}  by replacing the old versions of a library in the top-$i$ highest relevant labels by a newer version  if certain conditions are satisfied.  
As seen in Algorithm \ref{alg:Cache-based library reranking},
if a newer version, $l^{new}$, of a library is in the cache  and they have smaller $R(d,l^{new})$ values (line 3), i.e., $R(d,l^{new}) < R(d,l)$, \tool{} will set the relevance of the new label to be $R(d,l)$  (line 4) and  remove the old versions from consideration (line 5).

The time-aware adjustment has a \textbf{recency bias} and uses the cache to modify the  top-$i$ highest $R(d,l)$ values. 
The update function $f$ (used in Algorithm \ref{alg:Time-aware adjustment} on line 6) is formulated as:
\begin{equation}
\label{eq:update}
    f(R(d,l)) =\begin{cases}
    R(d,l) + \alpha \times \bar{R}&  l \in cache \\
    R(d,l)&  l \notin cache
    \end{cases}
\end{equation}
where the magnitude of $\alpha$ is determined by the relative recency of $l$ in the cache. 
More recently observed libraries are more likely to be the label of a vulnerability report. 
The parameter values require careful selection.
If $\alpha$ and $\bar{R}$ are too big, the adjustment function dominates the predictions of \tool{}.
Conversely, if they are too small, they do not affect the final scores. $\alpha$ and $\bar{R}$ are defined as follows:
\begin{align}
    \alpha &= \frac{M}{L_{recency}+1} \\
    \bar{R} &=\frac{\sum_{j=1}^{i}R(d,l_j)}{i}
\end{align}
where $M$ determines the magnitude of favouring recently vulnerable library versions.
$L_{recency}$ is the relative recency for label $l$ in the cache, which ranges from 0 to $c-1$. 0 implies that the label was just added, while a recency of $c-1$ implies that the label is the least recently used label in the cache.
$\bar{R}$ is the average of  top-$i$ $R(d,l)$ values.
The values of $M$ and $i$ are tuned on the validation dataset.





\section{Evaluation}
\label{sec:experiments} 


\subsection{Implementation details}
\label{subsec:implementation_details}
We implement \tool{} and the baseline approaches using the PyTorch library
and Python. The models are trained and evaluated on a Docker environment running Ubuntu 18.04
with Intel(R) i7-10700K @ 3.8GHz, 64GB RAM, and 2 NVIDIA RTX 2080 Ti GPU (11GB of graphics memory for each). For tuning LightXML's hyper-parameters, we run on AMD EPYC 7643 @ 2.3GHz, 512GB, and 4 RTX A5000. 

The detailed hyper-parameters of  LightXML and \tool{} are shown in Table~\ref{tab:lightxmlparams} and \ref{tab:chronosparams}, respectively. We run LightXML and \tool{} 5 times. 
These parameters are tuned through a grid search on the validation dataset considering the following possible values:
$c$ is in \{100, 200, 300, 400\}, $M$ is in \{0.5, 1, 2, 4, 8, 16, 32\}, $i$ is in \{5, 10, 50, 100\}, $x$ is in \{50, 60, 70, 80, 90, 95\}, and $y$ is in \{5, 6, 7, 8, 9, 10, 13, 15\}. 
The details of the grid search for LightXML are provided in the replication package \cite{implementation}. 
LightXML produced slightly different results with a standard deviation of 0.005, while \tool{} produced the same results each time. 
As the results are stable, we find that it is not necessary to repeat the experiments more than 5 times as our findings will not change even with more runs.

\begin{table}[t]
\centering
\caption{Parameters used for LightXML}
\label{tab:lightxmlparams}
\begin{tabular}{l|c}
\hline
\textbf{Parameter} & \textbf{Value} \\ 
\hline
Learning rate      & 0.0001        \\ 
Epoch              & 30             \\ 
Batch size         & 4              \\ 
SWA warmup         & 10             \\ 
SWA step           & 200            \\ 
Feature & Transformer generated vectors \\ 
\hline
\end{tabular}
\end{table}

\begin{table}[t]
\centering
\caption{Parameters used for Chronos}
\label{tab:chronosparams}
\begin{tabular}{l|c}
\hline
\textbf{Parameter} & \textbf{Value} \\ 
\hline
cache size ($c$)       & 300        \\ 
ranking-related factor ($M$)              & 8               \\ 
update range ($i$)       & 10              \\ 
$x$        & 50             \\ 
$y$         & 15            \\ 
\hline
\end{tabular}
\end{table}

\subsection{Dataset}
\label{subsec:dataset}
\begin{table}[]
\centering
\caption{The statistics of training, validation and testing dataset}
\label{tab:dataset}
\begin{tabular}{l|cc}
\hline
     \textbf{Dataset} & \textbf{\#Vulnerability Reports} & \textbf{\#Labels}\\ \hline
Training   & 3111   &  1378    \\ 
Validation & 1814   &  1094     \\ 
Testing    & 2740   &  1432     \\ \hline
\end{tabular}
\end{table}

To evaluate effectiveness of our approach, we use a dataset of 7,665 vulnerability reports with 4,682 labels from the NVD (National Vulnerability Database) and SCA (Software Composition Analysis) vulnerability database, initially collected by Chen et al \cite{chen2020automated}. 
Each report comprises a unique CVE ID, its vulnerability description, a list of web references, its CPE (Common Platform Enumeration) configuration,
and its labels (i.e., the affected libraries). 
For a fair comparison, we use the same preprocessing steps done by Haryono et al.~\cite{stefanus2022automated}. 
Each vulnerability report is a single document after applying these preprocessing steps:

\begin{itemize}
    \item \textbf{Description}: Non-alphanumeric characters and non-noun words are removed. Words that appear in more than 30\% of the vulnerability data are removed.
    \item \textbf{References}: Non-alphanumeric characters are replaced with whitespace.
    \item \textbf{CPE configuration}: Possible library names are retrieved using a regular expression based on the CPE format~\cite{buttner2007common}.

\end{itemize}

Finally, we have a dataset of 7,665 vulnerability reports with 2,817 labels. 
We split the dataset chronologically into training/validation/testing datasets. 
Our dataset comprises vulnerability reports published in a span of six years (2014-2019). 
The training/validation/testing splits follow the ratio 3:1:2. 
Particularly, vulnerability reports from years of 2014-2016, 2017 and 2018-2019 form the training, validation, and testing dataset  respectively.
Table \ref{tab:dataset} shows each dataset in detail.

\subsection{Experimental Metrics}
\label{subsec:metric}
Following previous works~\cite{chen2020automated,stefanus2022automated}, we evaluate the effectiveness of \tool{} and three baselines in terms of Precision (P), Recall (R) and F1-score (F1) calculated for the top-k prediction results with k=1,2,3. 
These metrics are standard metrics for the evaluation of XML tasks in prior studies \cite{stefanus2022automated, chen2020automated, jiang2021lightxml}. 
Particularly, for each technique, we obtain their prediction score for the possible labels of a given vulnerability report and then rank the labels based on the score to obtain the top-k prediction. 

Given a top-k prediction $lb\_k(v)$ and the
actual labels $\hat{lb}(v)$ for a given vulnerability report $v$, $P@k$
and $R@k$ are defined as follows:
\begin{equation*}
    \begin{array}{cc}
    P@k(v) = \dfrac{lb\_k(v) \cap \hat{lb}(v)} {\color{black}{min (k, |\hat{lb}(v)|)}}     & R@k(v) = \dfrac{lb\_k(v) \cap \hat{lb}(v)}{|\hat{lb}(v)|}  \\
    \end{array}
\end{equation*}

{\color{black}We normalize $P@k$ to compare each approach against an ideal approach. For example, in our dataset, 60.58\% of vulnerability reports are assigned only one label. 
For these reports, without normalization -- the min() expression in the denominator -- the maximum achievable P@3 is 0.33. 
The normalized P@k formula above considers this best possible score.
The results for original $P@k$ is presented in Appendix~\ref{sec:appendix}. }

Then, we compute the average of the precision and recall calculated above to obtain the $P@k$ and $R@k$ that we use to compare the performance between \tool{} and three baselines ($n$ refers to the number of labels):
\begin{equation*}
    \begin{array}{cc}
      P@k = \dfrac{1}{n} \sum_{v=1}^{n} P@k(v)   &  R@k = \dfrac{1}{n} \sum_{v=1}^{n} R@k(v)\\
    \end{array}
\end{equation*}
Finally, we compute $F1@k$, which is the harmonic mean of $P@k$ and $R@k$.
\begin{equation*}
F1@k =2 \times \frac{P@k \times R@k}{P@k + R@k}
\end{equation*}

\subsection{Baseline Approaches}
\label{subsec:baseline}
To assess \tool{}, we use the following baselines:

\noindent    \textbf{CPE Matcher}: CPE Matcher is a simple baseline proposed by Chen et al.~\cite{chen2020automated}. CPE matcher uses the libraries listed in the CPE configuration of a vulnerability report. Particularly, CPE matcher retrieves library names and versions from the CPE configurations and outputs them as the labels on the vulnerability report.
    
\noindent   
    \textbf{Traditional IR}. We use TF-IDF with bag-of-ngrams ($n \leq 2$) to obtain feature vectors for the vulnerability reports and the labels. For each report, the cosine similarity between its feature vector and every label is computed. The top ranked labels are selected as output.

\noindent    \textbf{Exact Matcher}: As simple approaches can sometimes outperform complex ones in software engineering tasks~\cite{fu2017easy,zeng2021deep,kang2022detecting}, we propose a handcrafted heuristic-based approach that we term an Exact Matcher, which directly matches labels to their occurrences in vulnerability reports. 
    Exact Matcher ranks the label based on the number of occurrences that the library name occurs in each vulnerability report and outputs the top-k labels that occurs most frequently. 

\noindent    \textbf{LightXML}: LightXML~\cite{jiang2021lightxml} is the best-performing XML technique on the library identification problem with non zero-shot setting in the experiments  by Haryono et al.~\cite{stefanus2022automated}. 
    LightXML is a deep learning-based XML technique that uses transformer-based models with dynamic negative sampling. Particularly, LightXML divides labels into clusters based on balance K-Means~\cite{malinen2014balanced} and represent vulnerability reports using dense 768-dimensional vectors obtained from transformer-based models such as RoBERTa~\cite{liu2019roberta}, BERT~\cite{kenton2019bert} and XLNet~\cite{yang2019xlnet}. 
    LightXML uses generative cooperative networks with dynamic negative label sampling to score all label clusters and returns possible libraries. 
    Finally, LightXML scores every returned label and outputs the top-k highest score labels. 
    


\subsection{Research Questions}
We aim to answer the following research questions:

\noindent \textbf{RQ1:} \textit{What is percentage of unseen libraries in practice?} This research question investigates the percentage of unseen libraries that do not belong to the training dataset  in practice. 
To answer this question, we investigate the percentage of seen and unseen libraries on our dataset as described in section  \ref{subsec:dataset}. Particularly, we count the number of seen and unseen libraries for each year from 2015 to 2019. 
We consider a library of a vulnerability report as an \textit{unseen label} if it does not appear in vulnerabilities from the training dataset, which includes vulnerabilities reports published chronologically before the reports in the testing dataset. 

\vspace{2mm}
\noindent \textbf{RQ2:} \textit{Is \tool{} effective in identifying libraries from vulnerability reports?} This research question concerns the ability of \tool{} in identifying libraries from vulnerability reports.  
To evaluate our approach, we evaluate \tool{} on a dataset of 7,665 real-world vulnerability reports in terms of Precision, Recall, and F1-score as described in section \ref{subsec:metric}. 
We compare our approach to multiple baselines, including the state-of-the-art technique, LightXML \cite{stefanus2022automated}, the CPE Matcher \cite{chen2020machine} and a handcrafted exact matching algorithm. We run each tool five times and report the average results.

\vspace{2mm}

\noindent \textbf{RQ3:} \textit{Which components of \tool{} contributes to its performance?}
\tool{} contains multiple components, including the data enhancement and time-aware adjustment.
In this research question, we investigate the contribution of each component in an ablation study.


\subsection{RQ1: Percentage of Unseen Labels per Year} 
\label{subsec:unseen_ratio}


\begin{table}[]
\centering
\caption{The statistics of seen and unseen libraries per year during the period 2015-2019}
\label{tab:unseen_ratio}
\begin{tabular}{c|c|cc}
\hline
     \textbf{Year} & \textbf{\#Total} & \textbf{\#Seen Libraries} & \textbf{\#Unseen Libraries} \\ \hline
2015 & 656         & 312  (47.6\%)       & 344 (52.4\%)           \\ 
2016 & 896         & 345  (38.5\%)       & 551 (61.5\%)         \\ 
2017 & 1094        & 329  (30.0\%)       & 725 (70.0\%)          \\ 
2018 & 1094        & 451  (41.2\%)       & 643 (58.8\%)          \\ 
2019 & 651         & 313  (46.5\%)       & 338 (53.5\%)         \\ \hline
\end{tabular}
\end{table}

\begin{table}[]
\centering
\caption{The statistics of vulnerability reports containing seen and unseen labels per year during the period 2015-2019. The \#Seen, \#FullUnseen and \#PartialUnseen denotes the number of vulnerability reports are related to only seen labels, only unseen labels and both seen and unseen ones, respectively.}
\label{tab:unseen_ratio_reports}
\begin{tabular}{c|c|ccc}
\hline
     \textbf{Year} & \textbf{\#Total} & \textbf{\#Seen} & \textbf{\#FullUnseen} & \textbf{\#PartialUnseen} \\ \hline
2015 & 981         & 551  (56.2\%)       & 292 (29.8\%)  &  430 (43.8\%)        \\ 
2016 & 1347         & 704  (52.3\%)       & 447 (33.8\%)  &  643 (47.7\%)        \\ 
2017 & 1814         & 896  (49.3\%)       & 837 (33.2\%)  &  918 (50.7\%)        \\ 
2018 & 1718         & 872  (49.5\%)       & 640 (46.1\%)  &  846 (50.5\%)        \\ 
2019 & 1022         &  498 (48.7\%)       & 458 (44.8\%)  &  525 (51.3\%)        \\ \hline

\end{tabular}
\end{table}

We investigate the percentage of seen and unseen labels in our dataset.
We count the number of seen and unseen labels and their associated vulnerability reports for each year from 2015 to 2019. 
As 2014 is the first year of our dataset, we exclude it from the table.
We consider a label unseen if it does not appear in previous years. The results are reported in Table \ref{tab:unseen_ratio} and \ref{tab:unseen_ratio_reports}. 

As shown in Table \ref{tab:unseen_ratio}, the percentage of unseen labels ranges from 52.4\% to 70\% during 2015-2019. 
In particular, unseen labels account for almost half of all vulnerable labels in every years, and the percentage of unseen labels is 70\% in 2017. 

Concerning the percentage of descriptions associated with unseen labels, Table \ref{tab:unseen_ratio_reports} shows that 43.8\% to 51.3\% vulnerability descriptions during 2015-2019 contain at least one unseen label. Moreover, there are up to 46.1\% vulnerability descriptions containing only unseen labels. 
These results reveal the limitations of a non zero-shot learning techniques on the problem as they cannot correctly predict unseen labels.

\begin{tcolorbox}
  \textbf{\underline{Answer to RQ1:} Up to 70\% of labels are unseen labels. This affects 43.8\% to 51.3\% of vulnerability descriptions per year. This suggests that existing approaches cannot correctly produce the right labels for half of all vulnerability reports each year.}  
  
  \end{tcolorbox}



\subsection{RQ2: Comparison with Baselines} 

\begin{table*}[t]
\centering
\caption{Comparison of the effectiveness of \tool{} with the state-of-the-art techniques. The bold numbers denote the best results for each metric.\tool{} w/o DE, \tool{} w/o TA denotes the results of \tool{} without Data Enhancement and Time-aware Adjustment, respectively.}
\label{tab:overall}
\begin{tabular}{l|lll|lll|lll|l}
\hline
\textbf{Model}          & \textbf{P@1}  & \textbf{R@1}  & \textbf{F1@1}  & \textbf{P@2}  & \textbf{R@2}  & \textbf{F1@2}  & \textbf{P@3}  & \textbf{R@3}  & \textbf{F1@3}  & \textbf{Avg. F1} \\  \hline
Exact Matching & 0.33 & 0.26 & 0.29 & 0.44 & 0.41 & 0.42 & 0.48 & 0.46 & 0.47 & 0.39       \\
CPE Matcher   & 0.27 & 0.26 & 0.26 & -    & -    & -    & -    & -    & -    & -          \\
Traditional IR & 0.34  & 0.25 	& 0.29 	& 0.36	& 0.33 	& 0.35 	& 0.41 	& 0.39 	& 0.40 	& 0.34 \\
LightXML        & 0.32 & 0.21 & 0.26 & 0.29 & 0.28 & 0.29 & 0.30 & 0.29 & 0.30 & 0.28     \\ \hline
ZestXML & 0.56 & 0.45 & 0.50 & 0.63 & 0.60 & 0.61 & 0.67 & 0.65 & 0.66 & 0.59       \\
\tool{}          & \textbf{0.75} & \textbf{0.61} & \textbf{0.67} & \textbf{0.80} & \textbf{0.75} & \textbf{0.77} & \textbf{0.82} & \textbf{0.79} & \textbf{0.80} & \textbf{0.75} \\ 
\tool{} w/o DE & 0.70 & 0.57 & 0.63 & 0.75    & 0.70    & 0.72   & 0.77    & 0.74   & 0.75    & 0.70          \\
\tool{} w/o TA & 0.60 & 0.49 & 0.54 & 0.70    & 0.67    & 0.68   & 0.73    & 0.71   & 0.72    & 0.65 \\ \hline
\end{tabular}
\end{table*}


We compare \tool{} against the baselines approaches with respect to Precision, Recall, and F1 at top-k predictions (k=1,2,3). 
The
detailed results are shown in Table \ref{tab:overall}. 

Table \ref{tab:overall} shows that \tool{} achieves an F1 of 0.75 on average with the F1@1 of 0.67, F1@2 of 0.77 and F1@3 of 0.80.
These results indicate that \tool{} consistently outperforms the baseline tools, outperforming the best baseline by 131.1\%, 83.3\%, 70.2\%, and 92.3\% in terms of F1@1, F1@2, F1@3, and average F1 respectively.  
Compared to LightXML, \tool{} outperforms it by {\color{black}167.9\%} in average F1.
Notably, LightXML underperforms the Exact Matching baseline in every metric. 
This highlights the challenge of the zero-shot experimental setting as LightXML was the best-performing approach in the experiments of the prior study~\cite{stefanus2022automated}

When considering only either Precision or Recall, \tool{} is still the best performing approach. 
On Precision, \tool{} is outperforms the best baseline by 127.3\%, 81.8\%, and 70.8\% in the top-1, top-2, and top-3 predictions respectively. 
On Recall, \tool{} outperforms the best baseline by 134.6\%, 82.9\% and 71.7\% in the top-1, top-2, and top-3 predictions respectively.  

Compared to ZestXML alone, \tool{} improves by 27\% in average F1. The improvements come from both increases in precision (up to 33.9\%) and recall (35.5\%). 
This highlights the contributions of the domain-specific components, i.e. data enhancement and time-aware adjustment.

\begin{tcolorbox}
  \textbf{\underline{Answer to RQ2:} Yes, \tool{} is 92.3\% better in average F1 compared to the strongest baseline. The improvements come from both increases in precision (up to 127.3\%) and recall (up to 134.6\%).}  
  
  \end{tcolorbox}

\subsection{RQ3: Ablation Study} 

In this experiment, we evaluate the relative contribution of two components, data enhancement and time-aware adjustment, to the overall performance of our approach, \tool{}. 
Table \ref{tab:overall} shows the results of our experiments. 

As shown in Table \ref{tab:overall}, removing each component reduces the overall performance of \tool{}.
The performance of \tool{} drops in every metric. 
The average F1 of \tool{} without data enhancement and time-aware adjustment are declined from 0.75 to 0.7 ( $\downarrow$ 6.7\%) and 0.65 ($\downarrow$ 9.2\%), respectively. 
This suggests that both data enhancement and the time-aware adjustment are crucial to the effectiveness of \tool{}.
Moreover, the results also suggest that the time-aware adjustment is more essential to \tool{}. 

\begin{tcolorbox}
  \textbf{\underline{Answer to RQ3:} All components of \tool{} contribute positively to its effectiveness. Without data enhancement and time-aware adjustment, the performance of \tool{} decreases by 6.7\% and 9.2\% in terms of average F1, respectively.}  
  \end{tcolorbox}

\section{Discussion}
\label{sec:discussion}

\subsection{Time Efficiency}

\begin{figure}
    \centering
    \includegraphics[width=0.6\linewidth]{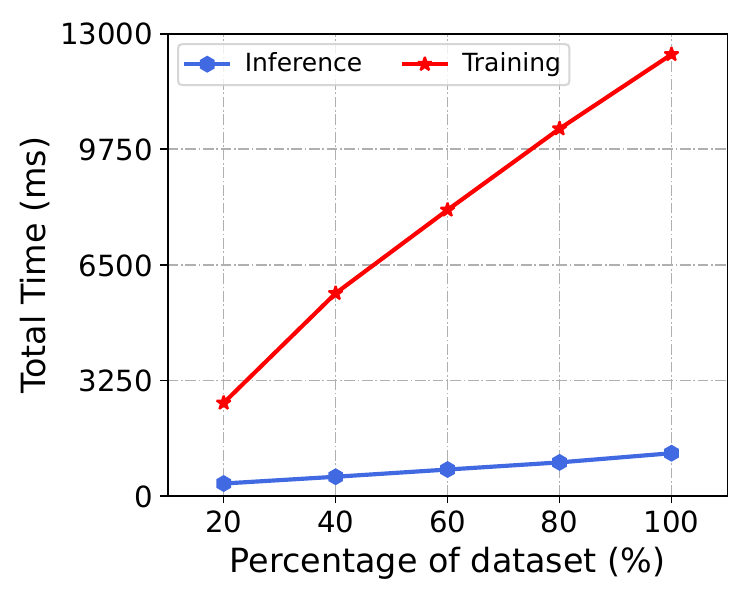}
    \caption{Training and inference time of \tool{} given different dataset sizes. The x-axis is the percentage of all vulnerability descriptions and the y-axis is the total time cost for training or inference. The total time of training or inference grows almost linearly with the size of dataset. It just costs just 1.62 milliseconds to train and 0.26 milliseconds for inference per vulnerability description. }
    \label{fig:time-efficency}
\end{figure}
For practical usage, \tool{} should work under a reasonable amount of time.
We investigate the efficiency of \tool{}.
We analyze the training time (the amount of time a model takes to learn all the training examples on average) and inference time (the amount of time a model takes to return all prediction results on average). 
The training time and inference time are related to two factors: the machine where the models run, and the size of  the dataset (i.e., how many vulnerability descriptions are used to train or to infer). 
We limit the models to only using 8 CPU cores to simulate running on a regular consumer-grade laptop.
As one would expect, a greater number of vulnerability descriptions takes a longer time to compute. 
There are 7,665 vulnerability descriptions in total.
We experiment with different dataset sizes. 
To reduce the effects of randomness, we repeat the experiments three times. 
The results are presented in Figure \ref{fig:time-efficency}.

The inference time is just a few thousands milliseconds and the training time is below fifteen thousand milliseconds with all vulnerability descriptions. 
The inference time and training time grow linearly with the size of the dataset, 
which shows the scalability of \tool{} in practice. 
\tool{} requires just 1.62 milliseconds for training and 0.26 milliseconds to infer labels for each vulnerability description.
This indicates that \tool{} is practical for use on a consumer-grade laptop.

\subsection{Qualitative Analysis}


\begin{figure}
    \centering
    \includegraphics[width=0.9\linewidth]{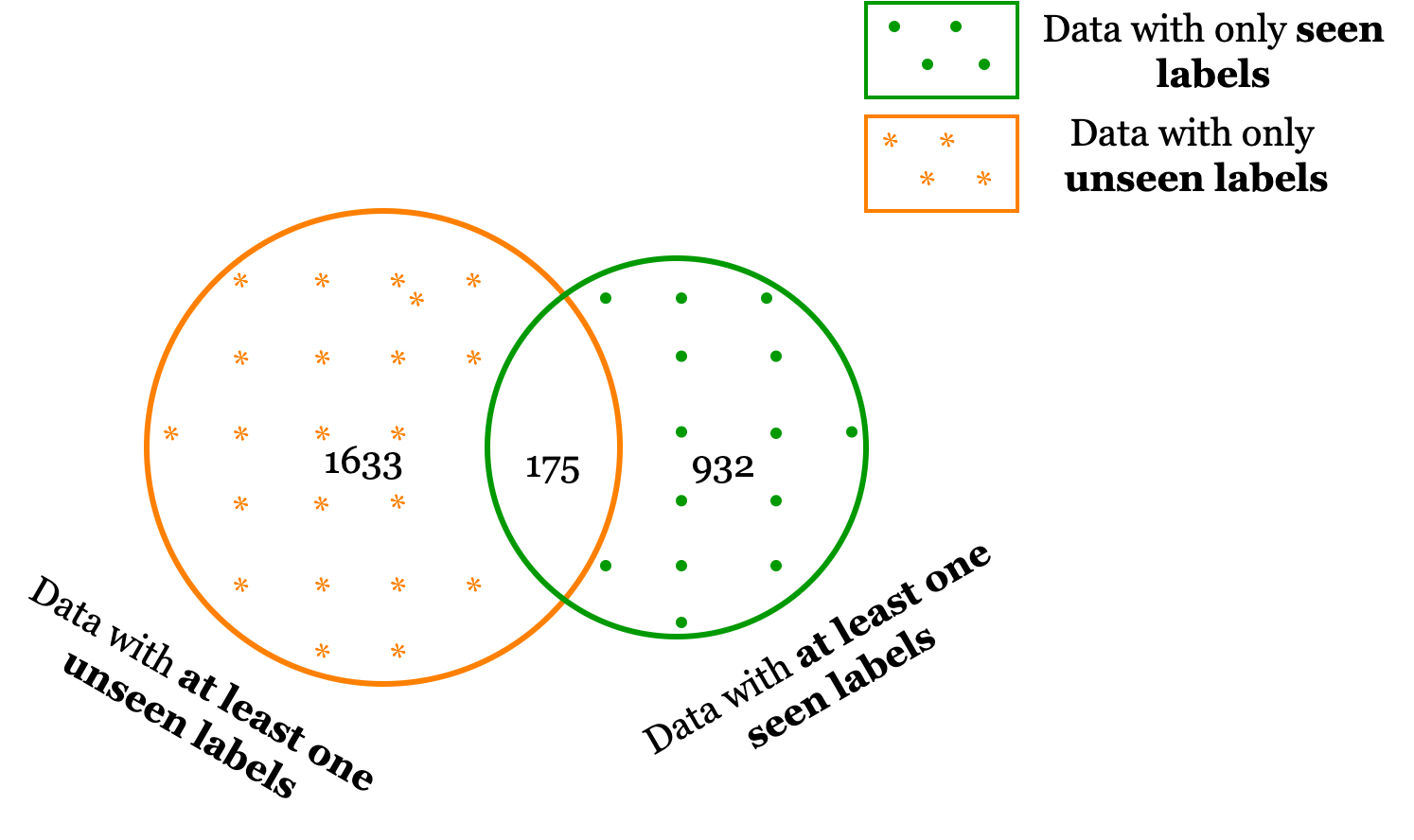}
    \caption{Distribution of vulnerability reports. We have a total of 2,740 test instances, of which 1,633 are reports with unseen labels and 932 are reports with seen labels. 175 reports have both seen and unseen labels.}
    \label{fig:venn}
\end{figure}

As we reformulate the problem as a zero-shot learning task, 
we investigate the performance of \tool{} in predicting labels without any training data. 
Figure \ref{fig:venn} shows the composition of the vulnerability reports in the testing dataset. 

\begin{table*}[t]
\centering
\caption{Comparison of \tool{} under different testing data. \tool{}\_FullSeen  and \tool{}\_PartialUnseen denotes the results of \tool{} on the vulnerability reports with only seen labels and at least one unseen labels, respectively}
\label{tab:data point}
\begin{tabular}{l|lll|lll|lll|l}
\hline
\textbf{Model}          & \textbf{P@1}  & \textbf{R@1}  & \textbf{F1@1}  & \textbf{P@2}  & \textbf{R@2}  & \textbf{F1@2}   & \textbf{P@3}  & \textbf{R@3}  & \textbf{F1@3}  & \textbf{Avg. F1} \\  \hline
\tool{}\_FullSeen   & 0.83 & 0.68 & 0.75 & 0.89 & 0.86 & 0.87 & 0.91 & 0.89 & 0.90 & 0.84      \\ 
\tool{}\_PartialUnseen & 0.70 & 0.58 & 0.64  & 0.76  & 0.69  & 0.72  & 0.77 & 0.73 & 0.75 & 0.70      \\
\tool{}         & 0.75 & 0.61 & 0.67 & 0.80  & 0.75 & 0.77 & 0.82 & 0.79 & 0.80 & 0.75       \\ \hline
\end{tabular}
\end{table*}

We report the result of our approach, \tool{}, on the data with only seen labels, some seen labels, and all data with respect to Precision, Recall, and F1 at top-k predictions (k=1,2,3). 
The detailed results are shown in Table \ref{tab:data point}.
\tool{} achieves an average F1 of 0.70, with an F1@1 of 0.64, F1@2 of 0.72, and F1@3 of 0.75 on the data with \textbf{some} seen labels. 
For the data with \textbf{only} seen labels, our approach achieves an average F1 of 0.84, with an F1@1 of 0.75, F1@2 of 0.87, and F1@3 of 0.9.
Comparing our \tool{}'s performance on data with only seen labels and data with some seen labels, \tool{} performs better on the data with only seen labels by 20\%.

This indicates that \tool{}'s performance on data with unseen labels still has room for improvement.
Nevertheless, \tool{} is able to perform reasonably well on the data with some seen labels.
We conclude that \tool{} achieves strong prediction results for the seen labels and is still effective at making good predictions on the unseen labels.

\begin{table}[t]
\centering
\caption{Comparison of the effectiveness of \tool{} with the state-of-the-art techniques in predicting unseen labels. \tool{} is able to correctly predict 694 previously unseen labels.}
\label{tab:unseen label predicting}
\begin{tabular}{c|cc}
\hline
\textbf{Model}  & \textbf{Success Rate}  & \textbf{\# Success Cases / Total} \\  \hline
LightXML & 0\%    & 0 / 957    \\
\tool{}  & \textbf{72.52\%}  & \textbf{694 / 957 }  \\  \hline
\end{tabular}
\end{table}


\begin{figure}
    \centering
    \includegraphics[width=\linewidth]{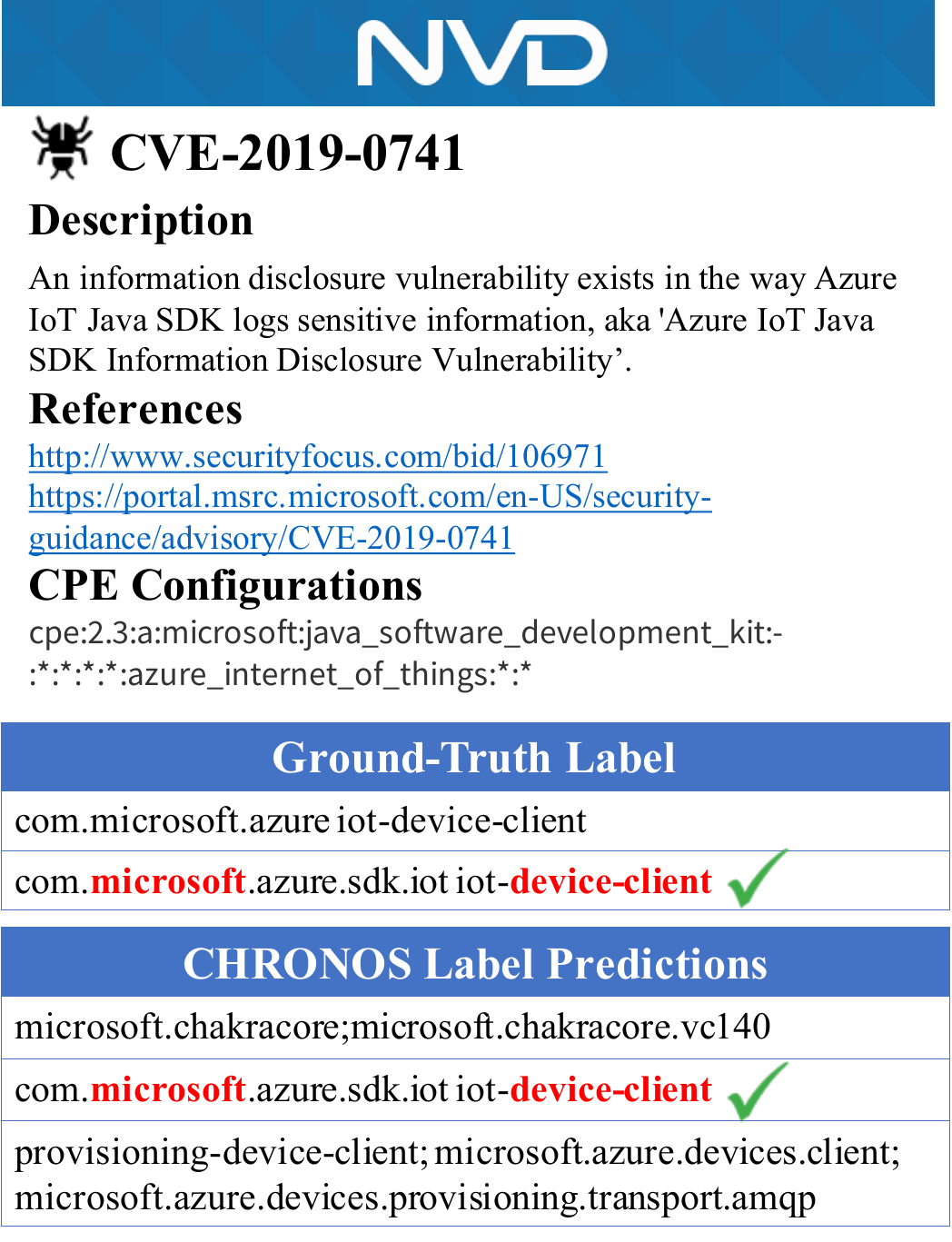}
    \caption{
    Unseen Label Example. NVD entry for CVE-2019-0741. The top of the image shows its NVD entry with description, references, and CPE configurations. The bottom of the image shows the ground-truth label and \tool{}'s prediction.}
    \label{fig:unseen label example}
\end{figure}

Table~\ref{tab:unseen label predicting} shows the improvements of \tool{} over the baseline LightXML in predicting unseen labels. 
While LightXML cannot predict unseen labels, our method successfully predicts at least one correct vulnerability for 72.52\% of them.
Figure~\ref{fig:unseen label example} shows an example of a vulnerability report where \tool{} successfully predicts an unseen label.
Given the text extracted from the vulnerability report, 
we compare the ground-truth label and  the predictions of \tool{}.
The red, bold characters in the ground-truth label and \tool{}'s predictions are text that do not appear in the vulnerability description. 
Even if the library name does not explicitly appear in the vulnerability descriptions, our method can predict them successfully.
This suggests that \tool{} successfully learns to 
identify relevant terms that are indicative of each library.

\subsection{Threats to Validity}

\textbf{Threats to internal validity} include possible errors in our implementation. A possible threat relates to the selection of \tool{}'s and baseline approaches' hyper-parameters. To mitigate this threat, we tune these hyper-parameters using grid search, selecting the best parameters using the validation dataset. 
While we tuned LightXML using a grid search with fewer parameters compared to \tool{} (learning rate,  batch size, and epoch),  
LightXML's effectiveness is limited as it is fundamentally unable to predict previously unseen labels.
We made the source code of our tool~\cite{implementation} and data~\cite{dataset} publicly available, allowing other researchers to validate our findings. 

\textbf{Threats to construct validity} 
are related to the suitability of our evaluation metrics. 
To minimize this threat, we have used the same performance metrics of precision, recall, and F1 of the top 3 predictions that were used in the previous studies~\cite{chen2020automated,stefanus2022automated}.
These are standard metrics used in the literature of XML approaches~\cite{jiang2021lightxml,prabhu2014fastxml}.

\textbf{Threats to external validity} are concerned with the generalizability of our experiments and findings. A possible threat is the dataset used in our experiments. We have utilized the same dataset from prior work~\cite{chen2020automated,stefanus2022automated}, containing vulnerabilities spanning over several years.
These data were collected and validated by security researchers in Veracode, hence, we believe that the threat is minimal. 
A final threat is related to our experiments on LightXML. 
If we consider an experimental setting where the data is provided as a stream, LightXML can be retrained after seeing each data point.
Unfortunately, LightXML requires 6 hours for training, which means that 6 (hours) x 2740 (entries) = 16,440 (hours) would be required for retraining LightXML. 
Therefore, a comparison of \tool{} and LightXML when data is provided as a stream is very expensive. 
We emphasize that this is a limitation of LightXML; LightXML will fail to predict previously unseen labels (occurring up to 70\% in Table~\ref{tab:unseen label predicting}) without frequent retraining, while \tool{} does not have this limitation.



\section{Related Work}

Software Composition Analysis is increasingly essential for securing software systems. In recent years, there have been many studies investigating the dependencies of software~\cite{han2020empirical, decan2019empirical, ponta2020detection, decan2018impact, zapata2018towards}. 
Many empirical studies have reported the impact of vulnerable dependencies in the software supply chain. For example, 
Decan et al. \cite{zapata2018towards} found that the number of vulnerability-affected packages in the npm network is growing over time, and half of the affected packages do not get fixed even when the fix is available. 
Lauinger et al. \cite{lauinger2018thou} also showed that around 37.8\% of the packages in the \texttt{npm} network have at least one vulnerable dependency. 
These findings demonstrated the growing importance of securing the software supply chain. 

Unfortunately, developers are slow in updating their vulnerable dependencies, leading to the risk of exploitation~\cite{kula2018developers, decan2021back}. 
Like our study, other researchers have recently proposed automated methods
that have emerged as a promising solution for speeding up the process~\cite{mirhosseini2017can, rombaut2022there}. 
For example, Mirhosseini et al.~\cite{mirhosseini2017can} found that projects that use automated pull requests upgrade dependencies 1.6x
often as projects that did not use any tools.
Other studies propose methods for detecting or obtaining more information about vulnerabilities from different software artifacts, including commits~\cite{nguyen2022hermes, zhou2021finding, nguyen2022vulcurator, zhou2021spi,sawadogo2022sspcatcher}, bug reports~\cite{zhou2017automated, chen2020machine,shu2021better}, and mailing lists~\cite{ramsauer2020sound, jovanovic2006pixy}.
Unlike these studies, we do not aim to detect vulnerabilities but to determine which libraries are vulnerable based on a vulnerability report.
Other methods help developers to check if a library vulnerability can be exploited~\cite{iannone2021toward,kang2022test} but already requires comprehensive information about the vulnerable library. 

For vulnerability reports, researchers have proposed methods to assist in the analysis of  vulnerabilities. 
Some studies focus on identifying affected versions~\cite{bao2022v} or predicting the exploitability of a vulnerability\cite{bozorgi2010beyond}. 
Other approaches use vulnerability reports to predict the key aspects, severity, or other properties of vulnerabilities\cite{guo2022detecting, bozorgi2010beyond, han2017learning, gong2019joint}. 
Another method models new attack techniques from textual descriptions of vulnerabilities \cite{binyamini2020automated}. 
Similarly, our work aims to make predictions of vulnerability reports.
However, we have a different goal of selecting libraries from a large space of possible labels.

Our study shows the importance of considering more practical experimental setups for analyzing vulnerability reports.
Other Software Engineering studies have also shown that overlooking time and other practical concerns may lead to brittle experimental results~\cite{tu2018careful, kang2022detecting, herzig2013s, christakis2016developers, sorensen2020norwegian, winter2022developers}. 
To address this practical challenge, our technique relies on a cache to leverage the time locality of the vulnerability reports. 
This phenomenon has been observed in other artifacts of software engineering. 
Tamrawi et al.\cite{tamrawi2011fuzzy} uses a caching strategy to enhance bug triaging by prioritizing developers who recently fixed related bugs. 
Caches of identifier names have been used  to improve language models for source code~\cite{hellendoorn2017deep,franks2015cacheca,tu2014localness}.

\section{Conclusion and Future Work}
\label{sec:conclusion}

Software Composition Analysis (SCA) depends on significant human effort in identifying every library that is affected by a vulnerability report.
Due to the large space of possible libraries, human effort can be error-prone.
Manual analysis relies on the human annotator's limited domain knowledge to match libraries against vulnerability reports that may not explicitly indicate every relevant library.
However, in this study, we show that the experimental setup considered in prior studies using extreme multi-label classification techniques may not consider a practical setting.

We reformulate the problem as a generalized zero-shot learning task, in which we face the challenge of predicting previously unseen labels.
Under the more realistic setting, prior approaches face a substantial drop in performance.

\tool{} uses zero-shot XML, data enhancement of the documents and labels, and time-aware adjustment of the labels.
\tool{} is frequently able to produce the right labels even if they were previously unseen.
\tool{} achieves an average F1 of 0.75, improving over the strongest approach identified in a prior study by {\color{black}167.9\%}.
Our experiments also indicate that each component of \tool{} contributes to its effectiveness.
These results suggest that the combination of techniques employed in \tool{} successfully addresses the challenge of predicting previously unseen libraries. 
Overall, the experiments suggest that \tool{} is effective for identifying libraries from vulnerability reports.

This study takes a large step forward in considering the real-world practical concerns of library identification from vulnerability reports.
Among other techniques in \tool{}, the use of reference data proved to be helpful in our experiments,
however, the references listed on NVD entries may be incomplete as the references were also identified through human analysis.
In the future, we will investigate methods of using software artifact traceability techniques~\cite{rodriguez2021leveraging,lin2021traceability} to link the NVD report to related artifacts (e.g. the commits on GitHub fixing the vulnerabilities).


\section{Data Availability}
\tool{}'s dataset and implementation are publicly available at \url{https://figshare.com/articles/software/Chronos-ICSE23/22082075} and \url{https://github.com/soarsmu/Chronos}, respectively. 

\section*{Acknowledgement}
This project is supported by the National Research Foundation, Singapore and National University of Singapore through its National Satellite of Excellence in Trustworthy Software Systems (NSOE-TSS) office under the Trustworthy Computing for Secure Smart Nation Grant (TCSSNG) award no. NSOE-TSS2020-02. Any opinions, findings and conclusions or recommendations expressed in this material are those of the author(s) and do not reflect the views of National Research Foundation, Singapore and National University of Singapore (including its National Satellite of Excellence in Trustworthy Software Systems (NSOE-TSS) office). 

Xuan-Bach D. Le is supported by the Australian Government through the Australian Research Council’s Discovery Early Career Researcher Award, project number DE220101057.

\balance
\bibliographystyle{IEEEtran}
\bibliography{IEEEabrv,zero_shot_vulnerable_library_identification.bib}

\begin{table*}[ht]
\centering
\caption{Comparison of the effectiveness of \tool{} with the state-of-the-art techniques. The bold numbers denote the best results for each metric.\tool{} w/o DE, \tool{} w/o TA denotes the results of \tool{} without Data Enhancement and Time-aware Adjustment, respectively. We use original P@k in this table.}
\label{tab:appendix_tab}
\begin{tabular}{l|lll|lll|lll|l}
\hline
\textbf{Model}          & \textbf{P@1}  & \textbf{R@1}  & \textbf{F1@1}  & \textbf{P@2}  & \textbf{R@2}  & \textbf{F1@2}  & \textbf{P@3}  & \textbf{R@3}  & \textbf{F1@3}  & \textbf{Avg. F1} \\  \hline
Exact Matching & 0.33 & 0.26 & 0.29 & 0.27 & 0.41 & 0.32 & 0.20 & 0.46 & 0.28 & 0.30       \\
CPE Matcher   & 0.27 & 0.26 & 0.26 & -    & -    & -    & -    & -    & -    & -          \\
Traditional IR & 0.34 	& 0.25 	& 0.29 	& 0.23 	& 0.33 	& 0.27 	& 0.19 	& 0.39 	& 0.25 	& 0.27 \\
LightXML        & 0.32 & 0.21 & 0.26 & 0.24 & 0.28 & 0.26 & 0.18 & 0.29 & 0.22 & 0.25     \\ \hline
ZestXML & 0.56 & 0.45 & 0.50 & 0.39 & 0.60 & 0.47 & 0.29 & 0.65 & 0.40 & 0.46      \\
\tool{}          & \textbf{0.75} & \textbf{0.61} & \textbf{0.67} & \textbf{0.49} & \textbf{0.75} & \textbf{0.60} & \textbf{0.36} & \textbf{0.79} & \textbf{0.49} & \textbf{0.59} \\ 
\tool{} w/o DE & 0.70 & 0.57 & 0.63   & 0.46  & 0.70 & 0.56   & 0.36  & 0.74 & 0.46  & 0.55     \\
\tool{} w/o TA & 0.60 & 0.49 & 0.54  & 0.43  & 0.67  & 0.52   & 0.32  & 0.71   & 0.44  & 0.50 \\ \hline
\end{tabular}
\end{table*}

\newpage

\appendix
\subsection{Evaluation Results using the Original Precision@K} \label{sec:appendix}

In this paper, we use a normalized version of $P@k$ to avoid the denominator being larger than the number of actual labels. In this appendix, we present the evaluation results on the original $P@k$, which is calculated as follows:

\begin{equation*}
    P@k(v) = \dfrac{lb\_k(v) \cap \hat{lb}(v)}{k} 
\end{equation*}

where $lb\_k(v)$ and $\hat{lb}(v)$ are top-k prediction and the
actual labels for a given vulnerability report $v$.

Table~\ref{tab:appendix_tab} illustrates that our primary conclusions and contributions remain consistent under the original $P@k$ setting. 
Specifically, \tool{} outperforms the standard zero-shot learning model in terms of Avg. F1 by 28\%, and it outperforms LightXML by 136\%.

\end{document}